\documentclass[conference]{IEEEtran}
\usepackage{booktabs} % For formal tables

\usepackage[utf8]{inputenc}
\usepackage{url}
\usepackage{xspace}
\usepackage{tikz}
\usetikzlibrary{arrows}
\usetikzlibrary{positioning}
\usepackage{float}
\usepackage{amsmath}
\renewcommand{\newcommand}{\providecommand}
\usepackage{ifthen}
\usepackage{url}
\usepackage{amssymb}
\usepackage{amsmath}
\usepackage{import}
\usepackage{xparse}
\usepackage{amsmath}
\usepackage{xspace}
\usepackage{datatool}
\usepackage{glossaries}

\providecommand\m[1]{\ensuremath{#1}\xspace}
\renewcommand{\m}[1]{\ensuremath{#1}\xspace}
\newcommand{\trval}[1]{\m{{\bf #1}}}

	\newcommand{\lrule}{\m{\leftarrow}}
	\newcommand{\cause}{\m{\stackrel{c}{\lrule}}}
	
	\newcommand{\ltrue}{\trval{t}}
	\newcommand{\lfalse}{\trval{f}}
	\newcommand{\lunkn}{\trval{u}}
	
	\newcommand{\Tr}{\ltrue}
	\newcommand{\Fa}{\lfalse}
	\newcommand{\Un}{\lunkn}

	\newcommand{\struct}{\m{I}}
	
	\newcommand{\I}{\m{\mathcal{I}}}

	\NewDocumentCommand\inter{g+g}{%
	  \IfNoValueTF{#1}
	    {\struct}
	    {\m{#1^{#2}}}}

	\newcommand{\ddd}{\m{\overline{d}}}

	\renewcommand{\int}{\m{\mathbb{Z}}}

	\newcommand{\leqp}{{\m{\,\leq_p\,}}}

	\NewDocumentCommand\subs{g+g}{%
	  \IfNoValueTF{#1}
	    {\m{/}}
	    {\m{#1/ #2}}}

	\newcommand{\logicname}[1]{\textsc{#1}\xspace}

	\newcommand{\idp}{\logicname{IDP}}

	\newcommand{\fodot}{\logicname{FO(\ensuremath{\cdot})}}
	
	\newcommand{\foid}{\logicname{FO(ID)}}

\newcommand{\ouracronym}[3]{%
	\newacronym{#1}{#2}{#3}
	\expandafter\newcommand\csname #1\endcsname{\gls{#1}\xspace}%
}
	\ouracronym{FO}{FO}{first-order logic}
	\ouracronym{MX}{MX}{Model Expansion}
	\ouracronym{MO}{MO}{Model Optimization}
	\ouracronym{ASP}{ASP}{Answer Set Programming}
	\ouracronym{TP}{TP}{Theorem Proving}
	\ouracronym{LP}{LP}{Logic Programming}
	\ouracronym{CP}{CP}{Constraint Programming}
	\ouracronym{FP}{FP}{Functional Programming}
	\ouracronym{KR}{KR}{Knowledge Representation}
	\ouracronym{CSP}{CSP}{Constraint Satisfaction Problem}
	\ouracronym{SMT}{SMT}{SAT Modulo Theories}
	\ouracronym{KBS}{KBS}{Knowledge Base System}
	\ouracronym{NNF}{NNF}{Negation Normal Form}
	\ouracronym{FNNF}{FNNF}{Flat Negation Normal Form}
	\ouracronym{DefNNF}{DefNNF}{Definition Negation Normal Form}
	\ouracronym{CDCL}{CDCL}{Conflict-Driven Clause-Learning}
	\ouracronym{LCG}{LCG}{Lazy Clause Generation}
	\ouracronym{AEL}{AEL}{Autoepistemic Logic}
	\ouracronym{OEL}{OEL}{Ordered Epistemic Logic}
	\ouracronym{AFT}{AFT}{Approximation Fixpoint Theory}

	\makeatletter
	\def\ifenv#1{
	\def\@tempa{#1}%
	\def\@ttempa{#1*}%
	\ifx\@tempa\@currenvir
	\expandafter\@firstoftwo
	\else
	\expandafter\@secondoftwo
	\fi
	}
	\makeatother

	\newcommand{\ddrule}[4]{\ensuremath{#1 \leftarrow #2 & \{#3\} & #4}}
	\newcommand{\drule}[2]{\ensuremath{#1 & \leftarrow & #2}}

	\newcommand{\darule}[4]{\ensuremath{#1 \leftarrow #2 & \{#3\} & #4}}
	\newcommand{\arule}[2]{\ensuremath{#1 \, &\leftarrow \, #2}}

	\newenvironment{ldef}{\left\{\begin{array}{l@{ \,}l@{\,}l}}{\end{array}\right\}}
	\newenvironment{ltheo}{\[\begin{array}{l}}{\end{array}\]\ignorespacesafterend}

	\newcommand{\LNDRule}[2]{
	\ifenv{array}
	{\drule{#1}{#2}}
	{ \ifenv{align}
		{\arule{#1}{#2}}
		{\ifenv{align*}
		{\arule{#1}{#2}}
		{ERROR: using LDRule in unsupported environment: \@currenvir}
		}
	}
	}

	\newcommand{\LDRule}[4]{
	\ifenv{array}
	{\ddrule{#1}{#2}{#3}{#4}}
	{ \ifenv{align}
		{\darule{#1}{#2}{#3}{#4}}
		{\ifenv{align*}
		{\darule{#1}{#2}{#3}{#4}}
		{ERROR: using LDRule in unsupported environment: \@currenvir}
		}
	}
	}

	\NewDocumentCommand\LRule{m+g+g+g}{%
		\IfNoValueTF{#2}%
		{#1.&}{%
		\IfNoValueTF{#3}
		{\LNDRule{#1}{#2.}}
		{\LDRule{#1}{#2.}{#3}{#4}}%
		}
	}

	\NewDocumentCommand\CLRule{m+g}{%
	\ifenv{array}
	{\cdrule{#1}{#2}}
	{ \ifenv{align}
		{\carule{#1}{#2}}
		{\ifenv{align*}
			{\carule{#1}{#2}}
			{ERROR: using CLRule in unsupported environment: \@currenvir}
		}
	}
	}

	\NewDocumentCommand\carule{m+g}{%
		\IfNoValueTF{#2}
			{\ensuremath{#1.}}
			{\ensuremath{#1 \, &\cause \, #2}}}
	\NewDocumentCommand\cdrule{m+g}{%
		\IfNoValueTF{#2}
			{\ensuremath{#1.}}
			{\ensuremath{#1 & \cause & #2}}}

	\newcommand{\algrule}[4]{
	\hbox{{#1}:}& 
	\quad #2 ~\longrightarrow~ #3 
	\hbox{~ if } #4\\
	}

	\newcommand{\AlgoRule}[4]{
	\ifenv{array}
	{\algrule{#1}{#2}{#3}{#4}}
		{ERROR: using AlgoRule in unsupported environment: \@currenvir}
	}

\newcommand{\commentstyle}{\color{Gray}}
	\usepackage[final]{listings}

	\lstdefinelanguage{idp}{
		morekeywords=[1]{query(}, %to ignore procedure calls to query
		morekeywords=[2]{namespace,vocabulary,theory,structure,procedure,term,set,formula, spec, specification,query},
		morekeywords=[3]{include,using,type,isa,contains,partial,extern,LFD,GFD,constructed,from,constraint,pred,supertype,of,subtype,define},
		morekeywords=[4]{int,float,char,string,nat},
		morekeywords=[5]{if,then,else,for,end},
		morecomment=[s]{/*}{*/},	
		morecomment=[l]{//}
	}
	\newcommand{\ignore}[1]{}

	\newboolean{nocomments}
	\setboolean{nocomments}{false}

	\newboolean{commentmargin}
	\setboolean{commentmargin}{true}

	\newcommand{\namedcomment}[3]{
		\ifthenelse{\boolean{nocomments}}
		{} %IF no comments, write nothing
		{ %Otherwise
			\ifthenelse{\boolean{commentmargin}}
				{ {\color{#3} \marginpar{\color{#3}\sc #2}#1}  } %Name in margin
				{  {\color{#3} {\sc #2}: #1}  } %Name not in margin
		}
	}
	\newcommand{\mnamedcomment}[3]{\ifthenelse{\boolean{nocomments}}{}{{\marginpar{ \color{#3}{\sc #2}:#1}}}}

	\usepackage{soul}

\usepackage{etoolbox}

\newcommand\setcitation[2]{%
  \csdef{mycommoncitation#1}{#2}}
\newcommand\getcitation[1]{%
  \csuse{mycommoncitation#1}}

\setcitation{IDP}{WarrenBook/DeCatBBD14}
\setcitation{idp}{WarrenBook/DeCatBBD14}
\setcitation{fodot}{tocl/DeneckerT08}
\setcitation{CP}{Apt03}
\setcitation{EZCSP}{lpnmr/Balduccini11}
\setcitation{KR}{Baral:2003}
\setcitation{ASPComp2}{lpnmr/DeneckerVBGT09}
\setcitation{ASPComp3}{journals/tplp/CalimeriIR14}
\setcitation{ASPComp4}{conf/lpnmr/AlvianoCCDDIKKOPPRRSSSWX13}
\setcitation{CPSupport}{ictai/DeCat13}
\setcitation{CPsupport}{ictai/DeCat13}
\setcitation{functionDetection}{iclp/DeCatB13}
\setcitation{fodot2asp}{DeneckerLTV12} %TODO replace by journal publication if one is publishedr
\setcitation{Tarskian}{DeneckerLTV12} %Same as the one above 
\setcitation{Inca}{iclp/DrescherW12}
\setcitation{csp2asp}{ijcai/DrescherW11}
\setcitation{DPLLT}{cav/GanzingerHNOT04}
\setcitation{AspInPractice}{synthesis/2012Gebser}
\setcitation{clasp}{ai/GebserKS12}
\setcitation{oclingo}{kr/GebserGKOSS12}
\setcitation{gringo}{lpnmr/GebserST07}
\setcitation{cmodels}{aaai/GiunchigliaLM04}
\setcitation{inputster}{tplp/Jansen13}
\setcitation{DLV}{tocl/LeonePFEGPS06}
\setcitation{LearningPaper}{TPLP/BruynoogheBBDDJLRDV} %TODO replace by published version
\setcitation{clog}{iclp/BogaertsVDV14} %TODO replace by journal publication if one is published
\setcitation{foc}{iclp/BogaertsVDV14}
\setcitation{inferenceClog}{ecai/BogaertsVDV14}
\setcitation{examplesClog}{nmr/BogaertsVDV14b} %TODO replace by better publication if possible
\setcitation{AFT}{DeneckerBV12}
\setcitation{KBS}{iclp/DeneckerV08}
\setcitation{KBPE}{inap/DePooterWD11}
\setcitation{lazyGrounding}{jair/CatDBS15} 
\setcitation{lazygrounding}{jair/CatDBS15} 
\setcitation{ASP}{marek99stable}
\setcitation{satid}{sat/MarienWDB08}
\setcitation{lazyclausegeneration}{constraints/OhrimenkoSC09}
\setcitation{FP}{ACMCS/Hudak89}
\setcitation{GroundingWithBounds}{jair/WittocxMD10}
\setcitation{SAT}{faia/SilvaLM09}
\setcitation{LTC}{iclp/Bogaerts14}
\setcitation{SPSAT}{ictai/DevriendtBMDD12}
\setcitation{BreakID}{cspsat/DevriendtBB14}
\setcitation{LCG}{stuckeyLCG}
\setcitation{MiniZinc}{conf/cp/NethercoteSBBDT07}
\setcitation{minizinc}{conf/cp/NethercoteSBBDT07}
\setcitation{amadini}{cpaior/AmadiniGM13}
\setcitation{bootstrapping}{lash/BogaertsJDJBD14}
\setcitation{GroundedFixpoints}{ai/BogaertsVD15}
\setcitation{LogicBlox}{datalog/GreenAK12}
\setcitation{proB}{journals/sttt/LeuschelB08}
\setcitation{NaturalInductions}{KR/DeneckerV14} %TODO replace by journal publication if one is published
\setcitation{LP}{jacm/EmdenK76}
\setcitation{SMT}{faia/BarrettSST09}
\setcitation{AF}{ai/Dung95}
\setcitation{ADF}{kr/BrewkaW10}
\setcitation{ADFRevisited}{ijcai/BrewkaSEWW13}
\setcitation{DefaultLogic}{ai/Reiter80}
\setcitation{DL}{ai/Reiter80}
\setcitation{AEL}{mo85}
\setcitation{minisat}{sat/EenS03}
\setcitation{completion}{adbt/Clark78}
\setcitation{wasp}{lpnmr/AlvianoDFLR13}
\setcitation{minisatid}{ictai/DeCat13}
\setcitation{lcg}{stuckeyLCG}
\setcitation{CEGAR}{jacm/ClarkeGJLV03}
\setcitation{cegar}{jacm/ClarkeGJLV03}
\setcitation{CuttingPlane}{or/DantzigFJ54}
\setcitation{kodkod}{tacas/TorlakJ07}
\setcitation{cdcl}{Marques-SilvaS99}
\setcitation{CDCL}{Marques-SilvaS99}
\setcitation{}{}
\setcitation{}{}
\setcitation{}{}
\setcitation{}{}
\setcitation{}{}
\setcitation{}{}
\setcitation{}{}
\setcitation{}{}
\setcitation{}{}
\setcitation{}{}
\setcitation{}{}
\setcitation{}{}
\setcitation{}{}
\setcitation{}{}
\setcitation{}{}
\setcitation{}{}
\setcitation{}{}
\setcitation{}{}
\setcitation{}{}
\setcitation{}{}
\setcitation{}{}
\setcitation{}{}
\setcitation{}{}
\setcitation{}{}
\setcitation{}{}
\setcitation{}{}
\setcitation{}{}
\setcitation{}{}
\setcitation{}{}
\setcitation{}{}
\setcitation{}{}
\setcitation{}{}
\setcitation{}{}
\setcitation{}{}
\setcitation{}{}
\setcitation{}{}
\setcitation{}{}
\setcitation{}{}
\setcitation{}{}
  
\newcommand\mycite[1]{%
      \ifcsname mycommoncitation#1\endcsname%
   \cite{\getcitation{#1}}%
  \else%
    \cite{#1}
  \fi%
}	
  
\usepackage{picins}
\usepackage{capt-of}

\usepackage{algorithm}
\usepackage[noend]{algorithmic}

\newcommand{\T}{T}
\newcommand{\pS}{\mathcal{S}}

\renewcommand{\I}{I}

\definecolor{Gray}{rgb}{0.3,0.3,0.3}

\newtheorem{definition}{Definition}
\newtheorem{example}{Example}

\ifCLASSINFOpdf
\else
\fi

\begin{document}
\title{Technical report of \\ ``The Knowledge Base Paradigm \\ Applied to Delegation Revocation''}

\author{\IEEEauthorblockN{Marcos Cramer}
\IEEEauthorblockA{TU Dresden}
\and
\IEEEauthorblockN{Zohreh Baniasadi}
\IEEEauthorblockA{University of Luxembourg}
\and
\IEEEauthorblockN{Pieter Van Hertum}
\IEEEauthorblockA{KU Leuven}}


%Telephone: (800) 555--1212, Fax: (888) 555--1212}


\maketitle

\begin{abstract}
In ownership-based access control frameworks with the possibility of delegating permissions and administrative rights, delegation chains will form. 
There are different ways to treat delegation chains when revoking rights, which give rise to different revocation schemes. 
In this paper, we investigate the problem of delegation revocation from the perspective of the \emph{knowledge base paradigm}.
A \emph{knowledge base} is a formal specification of domain knowledge in a rich formal language.
Multiple forms of inference can be applied to this formal specification in order to solve various problems and tasks that arise in the domain.
In other words, the paradigm proposes a strict separation of concerns between information and problem solving.
The knowledge base that we use in this paper specifies the effects of the various revocation schemes.
By applying different inferences to this knowledge base, we can solve the following tasks: to determine the state of the system after a certain delegation or revocation; to interactively simulate the progression of the system state through time; to determine whether a user has a certain permission or administrative right given a certain state of the system; to verify invariants of the system; and to determine which revocation schemes give rise to a certain specified set of desired outcomes.
\end{abstract}


\section{Introduction}
% \begin{itemize}
%  \item Delegation and revocation
%  \item Hagstr\"om's classification: The three dimensions.
%  \item The knowledge base paradigm.
%  \item How we apply the KB paradigm to delegation revocation.
% \end{itemize}

In ownership-based frameworks for access control, it is common to allow principals (users or processes) to grant both permissions and administrative rights to other principals in the system. Often it is desirable to grant a principal the right to further grant permissions and administrative rights to other principals. This may lead to delegation chains starting at a \emph{source of authority} (the owner of a resource) and passing on certain permissions to other principals in the chain \cite{Li,Tamassia,Chander,Yao}.

Furthermore, such frameworks commonly allow a principal to revoke a permission that she granted to another principal \cite{Hagstrom,Zhang,Chander,BarkerBoella}. Depending on the reasons for the revocation, different ways to treat the chain of principals whose permissions depended on the second principal's delegation rights can be desirable \cite{Hagstrom,Cramer}. For example, if one is revoking a permission given to an employee because he is moving to another position in the company, it makes sense to keep the permissions of principals who received them from this employee; but if one is revoking a permission from a user who has abused his rights and is hence distrusted by the user who granted the permission, it makes sense to delete the permissions of principals who received them from this user. Any algorithm that determines which permissions to keep intact and which permissions to delete when revoking a permission is called a \emph{revocation scheme}. %Revocation schemes are usually defined in a graph-theoretical way on the graph that represents which authorizations between the principals are intact.

Hagstr\"om et al.\ \cite{Hagstrom} have presented a framework for classifying possible revocation schemes along three different dimensions: the extent of the revocation to other grantees (propagation), the effect on other grants to the same grantee (dominance), and the permanence of the negation of rights (resilience). Since there are two options along each dimension, there are in total eight different revocation schemes in Hagstr\"om et al.'s framework. This classification was based on revocation schemes that had been implemented in database management systems \cite{Griffiths,Fagin,Bertino96,Bertino97}.

In this paper, we investigate the problem of delegation revocation from the perspective of the \emph{knowledge base paradigm}, a declarative programming paradigm based on the idea of strictly separating information and problem solving.
A \emph{knowledge base} is a formal specification of domain knowledge in a rich formal language. 
Multiple forms of inference can be applied to this formal specification in order to solve various problems and tasks that arise in the domain.
% In other words, the paradigm applies a strict separation of concerns between information and problem solving.
The \idp system is an implementation of the knowledge base paradigm with associated formal language \fodot, an extension of first-order logic \mycite{decat2018predicate}.

In this paper we present an application of the knowledge base paradigm to delegation revocation, realized in the \idp system.
We have written a formal specification of the eight revocation schemes in Hagstr\"om et al.'s framework, which formally described their effects.
By applying different inferences to this specification, we can solve various tasks that can be useful both for implementing a system which allows for these revocation schemes and for supporting a user of such a system: 
Given a certain state of the system and a certain action (delegation or revocation), one inference determines the state of the system after this action. 
Another inference can interactively simulate the progression of the system state through time.
A third determines whether a user has a certain permission or administrative right given a certain state of the system. 
A fourth inference allows to verify that the system satisfies certain invariants.
Finally, there is an inference that allows a user to specify a set of desired outcomes (e.g.\ that a certain user should no longer have a certain right while another user is unaffected) and determine which actions give rise to this outcome. 

This work constitutes a proof of concept, and we hope that it will inspire other researchers in computer security to consider the possibility of applying the methodology of the knowledge base paradigm to their research.

% \idp is a \emph{Knowledge Base System}, which combines a declarative specification in \fodot, with imperative management of the specification via the Lua scripting language. An \fodot specification theory consists of formulas in first-order logic and \emph{inductive definitions}. 
% Inductive definitions are essentially logic programs in which clause bodies can contain arbitrary first-order formulas.
% The combination of the declarative specification and the imperative programming environment makes this logic programming tool suitable for solving a large variety of different problems.
% 
% In this paper, we show that revocation schemes can be efficiently implemented in \idp by modelling them as \idp theories. One of the key features that make \idp a very efficient tool for implementing revocation schemes is the possibility to use inductive definitions for defining functions and predicates in \fodot, since the formal definition of revocation schemes can be captured in an elegant way as an inductive definition. 

The paper is structured as follows: We introduce Hagstr\"om et al.'s delegation revocation framework in Section \ref{sec:rs}. In Section \ref{sec:KB} we motivate and describe the knowledge base paradigm, the \idp system and its specification language \fodot. Section \ref{sec:KBrs} presents the application of the knowledge base paradigm to Hagstr\"om et al.'s framework. Section \ref{sec:related} discusses related work and section \ref{sec:conclusion} concludes.

\section{The delegation revocation framework}
\label{sec:rs}
% \begin{itemize}
%  \item Take over from DBSec paper, but don't explain each of the eight revocation schemes separately (takes too much space). (I suggest we just explain on revocation scheme here in detail though an example, and put the examples of the other seven revocation schemes into the appendix.)
% \end{itemize}

In this section we present Hagstr\"om et al.'s \cite{Hagstrom} delegation revocation framework. 
%Our presentation is based on the presentation in \cite{Aucher}.

Let $P$ be the set of principals (users or processes) in the system, let $O$ be the set of objects for which authorizations can be stated, and let $A$ be the set of access types, i.e.\ of actions that principals may perform on objects. For every object $o \in O$, there is a \emph{source of authority} (\emph{SOA}), for example the owner of file $o$, which is a principal that has full power over object $o$ and is the ultimate authority with respect to accesses to object $o$. For any $a \in A$ and $o \in O$, the SOA of $o$ can grant the right to access $a$ on object $o$ to other principals in the system, and can also delegate the right to grant access and to grant this delegation right. Additionally, the framework allows for negative authorizations, 
%Due to space limitations, we restrict ourselves to the fragment of Hagstr\"om et al.'s framework consisting of positive authorizations only.
which can be used to block a principal's access or delegation rights without deleting any authorization.

We assume that all authorizations in the system are stored in an 
authorization specification, and that every authorization is of the form 
$(i,j,a,o,\langle\mathit{sign}\rangle,b^1,b^2)$, where $i,j \in P$, $a \in A$, $o \in O$, 
$\langle\mathit{sign}\rangle$ is $+$ or $-$, 
and $b^1$ and $b^2$ are booleans ($T$ or $F$). The meaning of this authorization is that principal $i$ is granting some permission concerning access type $a$ on object $o$ to principal $j$. 
If $\langle\mathit{sign}\rangle$ is $+$, then the permission is a positive permission, else it is a negative permission. 
If $b^1$ is $T$, the permission contains the right to delegate the permission further, i.e.\ to issue positive permissions. If $b^2$ is $T$, the permission contains the right to issue negative permissions. 
Since it does not make sense to combine a negative permission with the right to delegate a positive or negative permission, $b^1$ and $b^2$ have to be both $F$ if $\langle\mathit{sign}\rangle$ is $-$.

% \parpic[r]{
\begin{table}
\label{fig:R}
% \vspace{1mm}
\begin{center}
\begin{tabular}{|c|}
% \hline
% $(T,T)R(T,T)$\\
% \hline
% $(T,T)R(T,F)$\\
% \hline
% $(T,T)R(F,T)$\\
% \hline
% $(T,T)R(F,F)$\\
% \hline
% $(T,F)R(T,F)$\\
% \hline
% $(T,F)R(F,F)$\\
% \hline
\hline
$(+,T,T)R(+,T,T)$\\
\hline
$(+,T,T)R(+,T,F)$\\
\hline
$(+,T,T)R(+,F,T)$\\
\hline
$(+,T,T)R(+,F,F)$\\
\hline
$(+,T,F)R(+,T,F)$\\
\hline
$(+,T,F)R(+,F,F)$\\
\hline
$(+,T,T)R(-,F,F)$\\
\hline
$(+,F,T)R(-,F,F)$\\
\hline
\end{tabular}
\end{center}
% \vspace{-3mm}
\captionof{table}{Definition of $R$}
\end{table}
% }

Hagstr\"om et al.\ define a \emph{permission} to be a triple $(\langle\mathit{sign}\rangle, b^1,b^2)$, where $\langle\mathit{sign}\rangle$ is $+$ or $-$, and 
$b^1$ and $b^2$ are booleans. They formalize the meaning of the two booleans $b^1$ and $b^2$ by defining a binary relation $R$ between permissions. $R$ is defined to hold precisely for those pairs of permissions shown in Table 1.\footnote{Note that in \cite{Hagstrom}, the definition of $R$ also includes the pair $(+,F,T)R(+,F,F)$ of permissions. The inclusion of this pair however contradicts the informal explanations that Hagst\"om et al.\ have given, so that we assume that it was included by mistake.} Informally $\pi R \rho$ means that a principal with permission $\pi$ can grant permission $\rho$ to other principals.

% \begin{table}[h]
% \label{fig:R}
% \begin{center}
% \begin{tabular}{|c|}
% \hline
% $(+,T,T)R(+,T,T)$\\
% \hline
% $(+,T,T)R(+,T,F)$\\
% \hline
% $(+,T,T)R(+,F,T)$\\
% \hline
% $(+,T,T)R(+,F,F)$\\
% \hline
% $(+,T,F)R(+,T,F)$\\
% \hline
% $(+,T,F)R(+,F,F)$\\
% \hline
% $(+,T,T)R(-,F,F)$\\
% \hline
% $(+,F,T)R(-,F,F)$\\
% \hline
% \end{tabular}
% \end{center}
% \caption{Definition of $R$}
% \end{table}
 
We say that a permission $\pi$ is \emph{stronger} than a permission $\rho$ iff the set of permissions that a principal with permission $\pi$ can grant is a strict superset of the set of permissions that a principal with permission $\rho$ can grant. So $(+,T,T)$ is stronger than any other permission, and all permissions other than $(+,F,F)$ and $(-,F,F)$ are stronger than both $(+,F,F)$ and $(-,F,F)$.\footnote{Note that Hagstr\"om et al.\ use the word ``stronger'' in a different sense, namely as a synonym for the relation $R$. This however is a rather misleading way to use the word ``stronger''. Hagstr\"om et al.\ do not define the notion that we call ``stronger'', but this notion is needed in order to formalize Hagstr\"om et al.'s system.}

% Due to space limitations, we restrict ourselves to the fragment of Hagstr\"om et al.'s framework consisting of positive authorizations only. Furthermore, n

Note that there is no interaction between the rights of principals concerning different access-object pairs $(a,o)$, so we can consider $a$ and $o$ to be fixed for the rest of the paper. For these two reasons, we simplify an authorization $(i,j,a,o,+,b^1,b^2)$ to $(i,j,+,b^1,b^2)$. %, and we simplify a permission $(+,b^1,b^2)$ to $(b^1,b^2)$.

\subsection{Delegation chains and connectivity property}
\label{delegation}
In this subsection we focus on the part of the system that does not involve negative authorizations. In Subsection \ref{negative} we will introduce negative authorizations.

Hagstr\"om et al.\ use the notion of a principal $p$ \emph{having a permission} $\pi$ without formally defining it. The intended meaning of this can be formalized using the notion of a \emph{rooted delegation chain}:\footnote{The idea to formalize Hagstr\"om et al.'s notion of \emph{having a permission} in this way was first proposed by Aucher et al.\ \cite{Aucher}.}

\begin{definition}
 A \emph{rooted delegation chain} for principal $i$ with respect to permission $\pi$ is a sequence $p_1,\dots,p_n$ of principals satisfying the following properties: 
% \vspace{-4mm}
 \begin{enumerate}
  \item $p_1$ is the source of authority.
  \item $p_n$ is $i$.
  \item For every integer $k$ such that $1 \leq k < n$, an authorization $(p_k,p_{k+1},+,b^1_k,b^2_k)$ is in place.
  \item For every integer $k$ such that $1 \leq k < n-1$, it holds that $(+,b^1_k,b^2_k) R (+,b^1_{k+1},b^2_{k+1})$.
  \item $(+,b^1_{n-1},b^2_{n-1})$ is stronger or equal to $\pi$.
%  \item \label{4} An authorization of the form $(p_{n-1},p_n,T,b)$ is in place, for some $b \in \{T,F\}$.
 \end{enumerate}
\end{definition}

%If the $b$ in item \ref{4} of this definition it $T$, then $i$ does not only have the right to perform action $a$ on object $o$, but also the right to delegate this right further. 

% A principal $j$ has delegation right if there is a rooted delegation chain ending in $j$. $j$ has access right if she has delegation right or if some principal with delegation right has granted her the access right, i.e.\ if there is a principal $i$ such that there is a rooted delegation chain for $i$ and the authorization $(i,j,+,F,T)$ or $(i,j,+,F,F)$ is in place.

\begin{definition}
 We say that a principal $p$ \emph{has permission} $\pi$ iff there is a rooted delegation chain for principal $p$ with respect to permission $\pi$.
\end{definition}

Hagstr\"om et al.'s framework allows an authorization of the form $(i,j,\langle\mathit{sign}\rangle,b^1,b^2)$ to be in the authorization specification only if $i$ has a permission that allows $i$ to grant this authorization. This is called the \emph{connectivity property}:

\vspace{4mm}
\noindent \textbf{Connectivity property:} \\
 \textit{For every authorization $(i,j,\langle\mathit{sign}\rangle,b^1,b^2)$ in the authorization specification, $i$ has a permission $\rho$ with $\rho R (\langle\mathit{sign}\rangle,b^1,b^2)$.}
\vspace{4mm}

This property can be viewed as an invariant that any system based on Hagstr\"om et al.'s framework needs to satisfy.

We visualize an authorization specification by a labelled directed graph as in the following example:

% \vspace{-3mm}
\begin{figure}[H]
\center
\begin{tikzpicture}[->,>=stealth',auto,node distance=4cm and 4cm,
                    main node/.style={rectangle,rounded corners=5,draw}]

  \node[main node] (A) at (0,1) {~A~~};
  \node[main node] (B) at (2,1) {~B~~};
  \node[main node] (C) at (4,1) {~C~~};
  \node[main node] (D) at (4,0) {~D~~};
  \node[main node] (E) at (6,0) {~E~~};

  \path[every node/.style={font=\sffamily\tiny},pos=0.5, sloped]
    (A) edge node [above=-.1] {$+, T, T$} (B)

    (B) edge node [above=-.1] {$+, T, F$} (C)

    (B) edge node [above=-.1] {$+, T, T$} (D);

\end{tikzpicture}
\caption{Authorization specification visualized as labelled directed graph}
\end{figure}
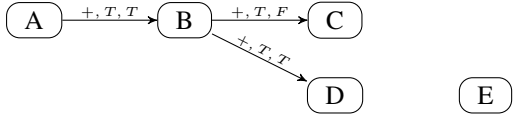
% \vspace{-3mm}
In this example, in which $A$ is the SOA (as in all forthcoming examples), the principals $A,B$ and $D$ have permission $(+,T,T)$, $C$ has permission $(+,T,F)$, and $E$ has no rights concerning the access type and object in question.

\subsection{Negative authorizations and inactivation of authorizations}
\label{negative}

A negative authorization from $i$ to $j$ can inactivate a positive authorization from $i$ to $j$ without deleting it. Hagstr\"om et al.\ \cite{Hagstrom} give two motivations for the use of negative authorizations: They can make a revocation resilient, i.e.\ make its effect time-persistent, and it can be used for temporarily taking away rights from a user without deleting anything from the authorization specification, so that it is easier to go back to the state that was in place before this temporary removal of rights. 
% The purpose of this is to make it possible to temporarily take away rights from a user without deleting anything from the authorization specification, so that it is easier to go back to the state that was in place before this temporary removal of rights. 

Hagstr\"om et al.\  leave it open whether negative permissions dominate positive ones or the other way round. If the system has a positive-takes-precedence conflict resolution policy, then positive permission take precedence; in this case, the resilient effect of a negative authorization is not given, so the only motivation to use negative authorizations in such a system it for a temporary removal of rights. If the system has a negative-takes-precedence conflict resolution policy, then negative permission take precedence; in this case, the distinction (explained below) between weak negative revocations and strong negative revocations disappears. In this paper, we assume the system to have a system has a positive-takes-precedence conflict resolution policy.

We assume the set of authorizations in the authorization specification to be divided into a set of \emph{active} authorizations and a set of \emph{inactive} authorizations. The revocation schemes that revoke a right by issuing a negative authorization inactivate some authorizations, i.e.\ turn some active authorizations into inactive authorizations. Negative authorizations are always active.

We define a rooted delegation chain to be active iff every authorization in it is active. We say that a principal $p$ \emph{actively has permission} $\pi$ iff there is an active rooted delegation chain for principal $p$ with respect to permission $\pi$. 
% \begin{definition}
%  An \emph{active rooted delegation chain} for principal $i$ is a chain $(p_1,\dots,p_n)$ of principals satisfying the following properties: 
%  \begin{enumerate}
%   \item $p_1$ is the source of authority.
%   \item $p_n=i$.
%   \item For every integer $k$ with $1 \leq k \leq n$, the authorization $(p_k,p_{k+1},T,T)$ is in place and the authorization $(p_k,p_{k+1},F,F)$ is not in place.
%  \end{enumerate}
% \end{definition}
% Now a positive authorization $(i,j,T,b)$ is considered inactive if there is no active rooted delegation chain for $i$. 

% When the authorization specification contains negative authorizations, the delegation right and access right definitions of section \ref{delegation} can no longer be applied as stated before, but must be modified by adding the word ``active'' to ``rooted delegation chain'': A principal $i$ has delegation right if there is an \emph{active} rooted delegation chain for $i$, and a principal $j$ has access right if there is a principal $i$ such that the authorization $(i,j,T,F)$ is in place and there is an \emph{active} rooted delegation chain for $i$.

Additionally to the connectivity property mentioned, we can also define an \emph{active connectivity property} as follows:

\vspace{4mm}
\noindent \textbf{Active-connectivity property:} \\
 \textit{For every active authorization $(i,j,\langle\mathit{sign}\rangle,b^1,b^2)$ in the authorization specification, $i$ actively has a permission $\rho$ with $\rho R (\langle\mathit{sign}\rangle,b^1,b^2)$.}
\vspace{4mm}

Unlike the connectivity property, the active-connectivity property is not mentioned by Hagstr\"om et al., but is actually also an invariant of the system they define.

\subsection{The three dimensions}
Hagstr\"om et al.\ \cite{Hagstrom} have introduced three dimensions according to which revocation schemes can be classified. These are called \emph{propagation}, \emph{dominance} and \emph{resilience}:

\textbf{Propagation.} 
The decision of a principal $i$ to revoke an authorization previously granted to a principal $j$ may either be intended to affect only the direct recipient $j$ or to affect all the other users in turn authorized by $j$. In the first case, we say that the revocation is \emph{local}, in the second case that it is \emph{global}. 

\textbf{Dominance.} 
This dimension deals with the case when a principal losing a permission in a revocation still has permissions from other grantors. 
%If these other grantors' delegation rights are independent of the revoker, there is nothing that the revoker can do about their grantings. However, if these other grantors have received their delegation rights through a path stemming from the revoker, 
If these other grantors' are \emph{dependent} on the revoker, she can dominate these grantors and revoke the permissions from them. This is called a \emph{strong} revocation. The revoker can also choose to make a \emph{weak} revocation, where permissions from other grantors to a principal losing a permission are kept. 

In order to formalize this dimension, we need to define what we mean by a principal's delegation rights to be \emph{independent} of another principal:
\begin{definition}
 A principal $j$ \emph{has delegation rights independent of} a principal $i$ with respect to permission $\pi$ iff there is an rooted delegation chain for $j$ with respect to $\pi$ that does not contain the principal $i$.
\end{definition}

\textbf{Resilience.} 
This dimension distinguishes revocation by removal of positive authorizations from revocation by negative authorizations which just inactivate positive authorizations. Given that we concentrate on the fragment of Hagstr\"om et al.'s framework without negative revocations, we will not explain this dimension in detail.
% We call revocations of the first kind \emph{deletes} and revocations of the second kind \emph{negatives}.

\subsection{The revocation schemes}
Since there are two options along each of the three dimensions, Hagstr\"om et al.\ defined eight different revocation schemes. %, of which this paper covers four. 
%For brevity, we just present five of the eight revocation schemes in detail, each with an example in which the authorization from $A$ to $B$ in the following authorisation specification is revoked according to the revocation scheme under consideration. 
Below we give semi-formal explanations of these eight revocation schemes. Fully formalized definitions are provided by the specification in Appendix A, which we describe in Subsection \ref{sec:spec}. We illustrate each revocation scheme with an example in which the authorization from $A$ to $B$ in the following authorisation specification is revoked according to the revocation scheme under consideration:  

% \vspace{-3mm}
\begin{figure}[H]
\center
\begin{tikzpicture}[->,>=stealth',auto,
                    main node/.style={rectangle,rounded corners=5,draw}]

  \node[main node] (A) at (0,1) {~A~~};
  \node[main node] (B) at (2,1) {~B~~};
  \node[main node] (C) at (4,1) {~C~~};
  \node[main node] (D) at (1,0) {~D~~};
  \node[main node] (E) at (4,0) {~E~~};
  \node[main node] (F) at (6,0) {~F~~};

  \path[every node/.style={font=\sffamily\tiny},pos=0.5, sloped]
    (A) edge node [above=-.1] {$+, T, T$} (B)
        edge node [above=-.1] {$+, T, T$} (D)

    (B) edge node [above=-.1] {$+, T, F$} (C)
        edge node [above=-.1] {$+, T, T$} (E)

    (D) edge node [above=-.1]  {$+, F, F$} (B)
        edge node [above=-.1] {$+, T, T$} (E)

    (E) edge node [above=-.1] {$+, F, F$} (F);

\end{tikzpicture}
\caption{Example authorization specification before revocation}
\end{figure}
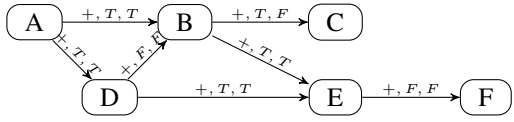
% \vspace{-7mm}

\subsubsection{Weak Local Delete}
A \emph{Weak Local Delete} (WLD) of a positive authorization from $i$ to $j$ has the following effect:
% \vspace{-1mm}
\begin{enumerate}
 \item The authorization from $i$ to $j$ is deleted.
 \item If step 1 causes $j$ to lose a permission, all authorizations emerging from $j$ that $j$ is no longer entitled to grant are deleted. 
 \item If $j$ still has another permission, each authorization deleted in step 2 is replaced by the strongest possible authorization that $j$ is entitled to grant and that is weaker than the deleted authorization. 
 \item For every authorization $(j,k,+,b^1,b^2)$ deleted in step 2, an authorization of the form $(i,k,+,b^1,b^2)$ is issued.
\end{enumerate}
% \vspace{-1mm}
Step 2 ensures that the connectivity property is satisfied at $j$. This being a local revocation scheme, step 4 ensures that all rights that users other than $j$ had before the operation are intact.

% \vspace{-5mm}
\begin{figure}[H]
\center
\begin{tikzpicture}[->,>=stealth',auto,node distance=4cm and 4cm,
                    main node/.style={rectangle,rounded corners=5,draw}]

  \node[main node] (A) at (0,1) {~A~~};
  \node[main node] (B) at (2,1) {~B~~};
  \node[main node] (C) at (4,1) {~C~~};
  \node[main node] (D) at (1,0) {~D~~};
  \node[main node] (E) at (4,0) {~E~~};
  \node[main node] (F) at (6,0) {~F~~};

  \path[every node/.style={font=\sffamily\tiny},pos=0.5, sloped]
    (A) edge [bend left=20, in=150, out=30, above=-.1] node {$+, T, F$} (C)
        edge node [above=-.1] {$+, T, T$} (D)
        edge [bend right, in=215,out=270] node [below] {$+, T, T$} (E)

%    (B) edge node [above=-.1] {$+, T, F$} (C)

    (D) edge node [above=-.1] {$+, F, F$} (B)
        edge node [above=-.1] {$+, T, T$} (E)

    (E) edge node [above=-.1] {$+, F, F$} (F);

\end{tikzpicture}
% \vspace{-2mm}
\caption{Weak Local Delete from $A$ to $B$}
\end{figure}
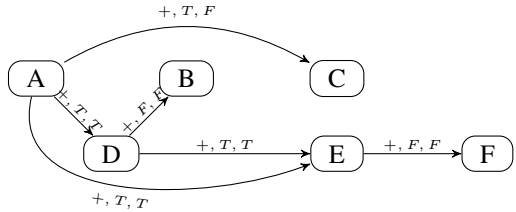
% \vspace{-8mm}

In this example, the authorizations from $B$ to $C$ and from $B$ to $E$ were deleted and not replaced by any authorization, because the permission $(+,F,F)$ that $B$ still has does no entitle $B$ to grant any authorization.

\subsubsection{Weak Global Delete}
A \emph{Weak Global Delete} (WGD) of a positive authorization from $i$ to $j$ has the following effect:
\begin{enumerate}
 \item The authorization from $i$ to $j$ is deleted.
 \item Recursively, any authorization emerging from a principal $k$ who loses a permission in step 1 or step 2 is deleted and replaced by the strongest possible authorization that $k$ is entitled to grant and that is weaker than the deleted authorization.
\end{enumerate}
The recursive step 2 ensures that the connectivity property is satisfied for the whole authorization specification after this operation.

% \vspace{-4mm}
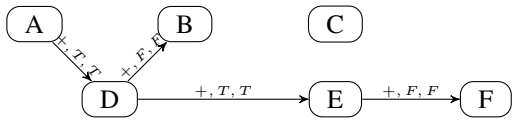
\begin{figure}[H]
\center
\begin{tikzpicture}[->,>=stealth',auto,node distance=4cm and 4cm,
                    main node/.style={rectangle,rounded corners=5,draw}]

  \node[main node] (A) at (0,1) {~A~~};
  \node[main node] (B) at (2,1) {~B~~};
  \node[main node] (C) at (4,1) {~C~~};
  \node[main node] (D) at (1,0) {~D~~};
  \node[main node] (E) at (4,0) {~E~~};
  \node[main node] (F) at (6,0) {~F~~};

  \path[every node/.style={font=\sffamily\tiny},pos=0.5, sloped]
    (A) edge node [above=-.1] {$+, T, T$} (D)

    (D) edge node [above=-.1] {$+, F, F$} (B)
        edge node [above=-.1] {$+, T, T$} (E)

    (E) edge node [above=-.1] {$+, F, F$} (F);

\end{tikzpicture}
\caption{Weak Global Delete from $A$ to $B$}
\end{figure}
% \vspace{-8mm}

\subsubsection{Strong Local Delete}
A \emph{Strong Local Delete} (SLD) of a positive authorization from $i$ to $j$ has the following effect:
% \vspace{-1mm}
\begin{enumerate}
 \item The authorization from $i$ to $j$ is deleted.
 \item Every authorization of the form $(k,j,+,b^1,b^2)$ such that $k$ is not independent of $i$ with respect to $(b^1,b^2)$ is deleted.
 \item If steps 1 and 2 cause $j$ to lose a permission, all authorizations emerging from $j$ that $j$ is no longer entitled to grant are deleted. 
 \item If $j$ still has another permission, each authorization deleted in step 3 is replaced by the strongest possible authorization that $j$ is entitled to grant and that is weaker than the deleted authorization. 
 \item For every authorization $(j,k,+,b^1,b^2)$ deleted in step 3, an authorization of the form $(i,k,+,b^1,b^2)$ is issued.
\end{enumerate}
% \vspace{-1mm}
The only difference to the Weak Local Delete is step 2, which is the step that makes this a strong revocation scheme.

% \vspace{-5mm}
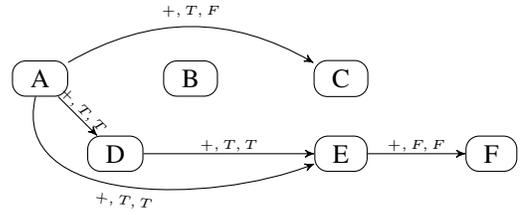
\begin{figure}[H]
\center
\begin{tikzpicture}[->,>=stealth',auto,node distance=4cm and 4cm,
                    main node/.style={rectangle,rounded corners=5,draw}]

  \node[main node] (A) at (0,1) {~A~~};
  \node[main node] (B) at (2,1) {~B~~};
  \node[main node] (C) at (4,1) {~C~~};
  \node[main node] (D) at (1,0) {~D~~};
  \node[main node] (E) at (4,0) {~E~~};
  \node[main node] (F) at (6,0) {~F~~};

  \path[every node/.style={font=\sffamily\tiny},pos=0.5, sloped]
    (A) edge [bend left=20, in=150, out=30] node {$+, T, F$} (C)
        edge node [above=-.1] {$+, T, T$} (D)
        edge [bend right, in=215,out=270] node [below] {$+, T, T$} (E)

 %   (B) edge node [above=-.1] {$T, F$} (C)

    (D) edge node [above=-.1] {$+, T, T$} (E)

    (E) edge node [above=-.1] {$+, F, F$} (F);

\end{tikzpicture}
% \vspace{-2mm}
\caption{Strong Local Delete from $A$ to $B$}
\end{figure}
% \vspace{-8mm}

\subsubsection{Strong Global Delete}
A \emph{Strong Global Delete} (SGD) of a positive authorization from $i$ to $j$ has the following effect:
% \vspace{-1mm}
\begin{enumerate}
 \item The authorization from $i$ to $j$ is deleted.
 \item Recursively, do the following
 \begin{enumerate}
  \item For any principal $k$ who has lost a permission in step 1, step 2.(a) or step 2.(c), all authorizations emerging from $k$ that $k$ is no longer entitled to grant are deleted.
  \item If $k$ still has another permission, each authorization deleted in step 2.(a) is replaced by the strongest possible authorization that $k$ is entitled to grant and that is weaker than the deleted authorization. 
  \item Any authorization of the form $(l,m,+,b^1,b^2)$, where $m$ is a principal who loses a permission in step 1, step 2.(a) or step 2.(c) and $l$ is not independent of $i$ with respect to $(b^1,b^2)$, is deleted.
 \end{enumerate}
\end{enumerate}
% \vspace{-1mm}
Here the recursive deletion procedure contains two different kinds of deletions: 2.(a) makes it a global revocation scheme and 2.(c) makes it a strong revocation scheme.

% \vspace{-4mm}
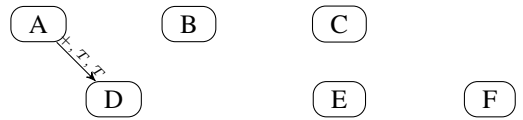
\begin{figure}[H]
\center
\begin{tikzpicture}[->,>=stealth',auto,node distance=4cm and 4cm,
                    main node/.style={rectangle,rounded corners=5,draw}]

  \node[main node] (A) at (0,1) {~A~~};
  \node[main node] (B) at (2,1) {~B~~};
  \node[main node] (C) at (4,1) {~C~~};
  \node[main node] (D) at (1,0) {~D~~};
  \node[main node] (E) at (4,0) {~E~~};
  \node[main node] (F) at (6,0) {~F~~};

  \path[every node/.style={font=\sffamily\tiny},pos=0.5, sloped]
    (A) edge node [above=-.1] {$+, T, T$} (D);

\end{tikzpicture}
% \vspace{-1mm}
\caption{Strong Global Delete from $A$ to $B$}
\end{figure}
% \vspace{-7mm}

\subsubsection{Negative revocations}
The negative revocations are similar to the positive revocations, only that instead of deleting positive authorizations, we inactivate them by issuing negative authorizations. We show this on the example of the Weak Local Negative. The other three negative revocation schemes are adapted versions of the corresponding delete revocations in a similar way. 

A \emph{Weak Local Negative} (WLN) of a positive authorization from $i$ to $j$ has the following effect:
% \vspace{-1mm}
\begin{enumerate}
 \item The negative authorization $(i,j,-,F,F)$ is added to the authorization specification.
 \item Any positive authorization from $i$ to $j$ is inactivated.
 \item If step 2 causes a permission of $j$ to ne inactivated, all authorizations emerging from $j$ that $j$ is no longer entitled to grant are inactivated. 
 \item If $j$ still actively has another permission, each authorization deleted in step 3 is replaced by the strongest possible authorization that $j$ is entitled to grant and that is weaker than the inactivated authorization. 
 \item For every authorization $(j,k,b^1,b^2)$ inactivated by step 3, an authorization of the form $(i,k,b^1,b^2)$ is issued.
\end{enumerate}
% \vspace{-1mm}
Unlike in the Weak Local Delete, we do not delete any authorizations, but just inactivate them. We graphically represent inactivated authoriazations by dashed lines.
% \vspace{-3mm}
\begin{figure}[H]
\center
\begin{tikzpicture}[->,>=stealth',auto,node distance=4cm and 4cm,
                    main node/.style={rectangle,rounded corners=5,draw}]

  \node[main node] (A) at (0,1) {~A~~};
  \node[main node] (B) at (2,1) {~B~~};
  \node[main node] (C) at (4,1) {~C~~};
  \node[main node] (D) at (1,0) {~D~~};
  \node[main node] (E) at (4,0) {~E~~};
  \node[main node] (F) at (6,0) {~F~~};

  \path[every node/.style={font=\sffamily\tiny},pos=0.5, sloped]
    (A) edge [bend left=20, in=160, out=45] node [above=-.1] {$+,T, F$} (C)
        edge [bend left=20] node [above=-.1] {$-,F, F$} (B)
        edge [dashed,bend right=10] node [above=-.1] {$+,T, T$} (B)
        edge  node [below=-.1] [above=-.1] {$+,T, T$} (D)
        edge [bend right, in=215,out=270] node [below] {$+,T, T$} (E)

    (B) edge [dashed] node [above=-.1] {$+,T, F$} (C)
        edge [dashed] node [above=-.1] {$+,T, T$} (E)

    (D) edge node [above=-.1]  {$+,F, F$} (B)
        edge node [above=-.1] {$+,T, T$} (E)

    (E) edge node [above=-.1] {$+,F, F$} (F);

\end{tikzpicture}
\caption{Weak Local Negative from $A$ to $B$}
\end{figure}
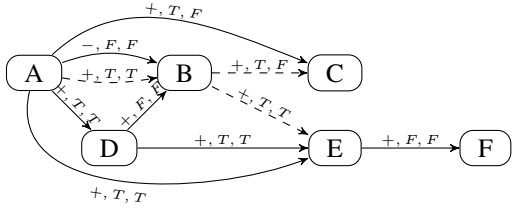

\section{The knowledge base paradigm and \\ the \fodot KB project}
\label{sec:KB}
% \subsection{The knowledge base paradigm}
Declarative systems have proven their merit in many application domains: From planning to scheduling to security contexts, many challenges have been tackled using declarative approach. For example, Barker et al.\ \cite{sacmat/Barker2009} used a rule-based approach to determine access rights in an access control system. Advantages of such an approach are readability and maintainability of a specification. However, only one task is supported in a rule based system, or in any other decalarative system. Every system has its own syntactical style, terminology, conceptualization, and designated style of inference (rule based systems do chaining, databases do querying, answer set programming generates answer sets, etc.). Yet, in all of them, propositions need to be expressed. Take, e.g., “each lecture takes place at some
time slot”. This proposition could be an expression to be deduced from a formal
specification if the task was a verification problem, or to be queried in a database,
or it could be a constraint for a scheduling problem. It is, in the first place, just a
piece of information and we see no reason why depending on the task to be solved, it should be expressed in a different formalism.

The \emph{knowledge base} (\emph{KB}) \emph{paradigm} \mycite{KBS} was proposed
as an answer to this. The KB paradigm applies a strict separation of concerns
to information and problem solving. A KB system allows information to be stored
in a knowledge base, and provides a range of inference methods. With these inference
methods various types of problems and tasks can be solved using the same
knowledge base. As such the knowledge base is neither a program nor a description
of a problem, it cannot be executed or run. It is nothing but information. However,
this information can be used to solve multiple sorts of problems. Many declarative
problem solving paradigms are mono-inferential: they are based on one form of inference.
In comparison, the KB paradigm is multi-inferential. 

The $\fodot$ KB project is a research project in which an implementation of the KB paradigm is being developed.  Its aim is to integrate different useful language constructs
and forms of inference from different declarative paradigms in one rich declarative
language and a KB system. So far, it has led to the KB language $\fodot$ \mycite{fodot} and the KB system \idp~\mycite{decat2018predicate}, which are used in this paper.

\subsection{The specification language \fodot}
\fodot refers to a class of extensions of first-order logic (FO). The language of the current version of the
IDP system (IDP 3) is FO(T, ID, Agg, arit, PF): FO extended with types, inductive definitions, aggregates, arithmetic
and partial functions (see \cite{tocl/DeneckerT08,tplp/PelovDB07}). In this work, we will only work (and as such, introduce) a subset of this language: $FO(T,ID)$: typed first-order logic with inductive definitions. Abusing notation, we will use $\fodot$ as an abbreviation of this language.

\subsection{An \fodot specification}
A specification of domain knowledge in \fodot can consist of 3 types of building blocks: a vocabulary $\Sigma$, a theory $\T$ and a (partial) structure $\pS$.
\paragraph{Vocabulary $\Sigma$.}
A \emph{vocabulary} is a declaration of the symbols used in the associated theories and structures.
It is a set $\Sigma$ of type symbols (denoted as $\Sigma_T$) and predicate symbols (denoted as $\Sigma_P$).
Every predicate $P$ of arity $n$ has a fixed type $[\tau_1,\ldots,\tau_n]$, where $\tau_1,\dots,\tau_n$ are type symbols.
Variables, atoms and first-order formulas are defined as usual. 

\paragraph{Theory $\T$.}
A \emph{theory} is a set of first-order formulas and inductive definitions.
An inductive definition $\Delta$ in $\fodot$ is a set of rules $\delta$ of the form $P(\overline{t})\leftarrow \varphi$, with $\varphi$ a first-order formula. 
We call $P(\bar t)$ the head ($head(\delta)$) and $\varphi$ the body ($body(\delta)$) of the rule.
The symbols that occur in the head of a rule $\delta$ in $\Delta$ are called the defined symbols in $\Delta$. All other symbols that occur in $\Delta$ are called the parameters of $\Delta$.
The semantics used for inductive definitions is the well-founded semantics; as argued in \cite{tocl/DeneckerT08}, this captures the intended meaning of all forms of inductive definitions commonly used in mathematics and computer science. 
A structure $\mathcal{S}$ satisfies $\Delta$ if the interpretation of a defined predicate $P$ in the well-founded model of $\mathcal{S}$, constructed relative to the restriction of $\mathcal{S}$ to the parameters of $\Delta$, is exactly the relation $P^{\mathcal{S}}$. 

The following example illustrates the use of an inductive definition in a theory by presenting the defintion of ``reachable'' in $\fodot$.
% A good example of a definition is a definition of the reachability relation in a graph.
\begin{example}
 Assume a vocabulary containing a type $Node$, and two predicates: $Edge(Node, Node)$ and $Reachable(Node,Node)$. Informally, $Edge$ states that there is an edge between two nodes, while $Reachable$ states that there is a path of edges between two nodes. We define what reachability means in terms of the edges, using an inductive definition $\Delta$ in $\fodot$:
 \begin{ltheo}
 \begin{ldef}
  \forall x: \LRule{Reachable(x,x)}\\
  \forall x\ y: \LRule{Reachable(x,y)}{\parbox[t]{0.2\textwidth}{$\exists z: Reachable(x,z) \; \land$ \\ 
  $Edge(z,y).$}}
 \end{ldef}  
 \end{ltheo}

\end{example}

\paragraph{Structure $\pS$.}
Given a vocabulary $\Sigma$, a \emph{partial structure} gives an interpretation to (a subset of) the elements of $\Sigma$.
Before we define formally what an interpretation is, we define the concept of a \emph{partial set}, which is a generalisation of a set in a 3-valued context:
A \emph{partial set} on domain $D$ is a function from $D$ to $\{\Tr,\Un,\Fa\}$. 
A partial set is two-valued (or total) if $\Un$ does not belong to its range. 
A \textit{(partial) structure} $\pS$ consists of a domain $D_\tau$ for all types $\tau$ in $\Sigma_T$ and an assignment of a partial set $P^\pS$ to each predicate or function symbol $P\in\Sigma_P$, called the \emph{interpretation} of $P$ in $\pS$.  
The interpretation $P^\pS$ of a predicate symbol $P$ with type $[\tau_1,\ldots,\tau_n]$ in $\pS$ is a
partial set on domain $D_{\tau_1}\times \ldots \times D_{\tau_n}$. 

In case the interpretation of a predicate $P$ in $\pS$ is a
two-valued set, we abuse notation and use $P^\pS$ as shorthand for
$\{\ddd|P^\pS(\ddd)=\Tr\}$. 

We call a partial structure \emph{total} if the interpretation $P^\pS$ of every predicate symbol $P\in\Sigma_P$ is a total set. Note that with the abuse of notation just explained, a total structure as we have defined it can be identified with a first-order structure as it is usually defined.

Given two partial structures $\pS = (D,\mathcal{I})$ and $\pS'=(D,\mathcal{I}')$, we write $\pS \leqp \pS'$ (and say $S$ \emph{is more precise than} $S'$, or $S'$ \emph{expands} $S$) iff 
%for every function symbol $f$, $f^{\mathcal{I}'} = f^\mathcal{I}$, and 
for every predicate symbol $P\in\Sigma_P$ with type $[\tau_1,\ldots,\tau_n]$ and every tuple $\bar d \in D_{\tau_1}\times \ldots \times D_{\tau_n}$ such that $P^\mathcal{I}(\bar d) \neq \Un$, we have $P^{\mathcal{I}'}(\bar d) = P^\mathcal{I}(\bar d)$.

\subsection{The reasoning engine}
In the $\fodot$ KB project, a implementation of a KB System was developed: the \idp system \mycite{decat2018predicate}. 
\idp takes an $\fodot$ specification (that is, a combination of vocabularies, theories and/or structures) and can do a number of reasoning tasks, by applying a suitable form of inference on this specification. 
Below, we present the inferences that we need in this paper:
\begin{itemize}
   \item \textbf{Modelexpand($\T,\pS$):} Input:  theory ${\T}$ and partial structure $\pS$.
     Output: either a total structure $\I$ such that $\I$ is a model of $\T$ and $\pS\leqp\I$, or \textit{UNSAT} if there is no such $\I$. 
     Modelexpand~\cite{lash08/WittocxMD08} is a generalization  for $\fodot$ theories of the modelexpansion task as defined in Mitchell et al.~\cite{MitchellT05}.
   \item \textbf{Allmodels($\T,\pS$):} Input:  theory ${\T}$ and partial structure $\pS$.
     Output: the set of all total structures $\I$ such that $\I$ is a model of $\T$ and $\pS\leqp\I$.
    % Complexity  of deciding the existence of a modelexpansion is in \textbf{NP}. 
%    \item[Modelcheck($\T,\S$):] Input: a total structure $\S$ and theory $\T$ over the vocabulary interpreted by $\S$. Output is the boolean value $\S \models \T$. 
%      Note that Modelcheck is a degenerate case of the Modelexpand inference, with $\pS$ a total structure.  %Complexity is in \textbf{P}.
%   \item[Minimize($\T,\pS,t$):] input: a theory $\T$, a partial structure
%     $\pS$ and a term $t$ of numerical type; output: a model $\I\geqp \pS$
%    of $\T$ such that the value $t^{\I}$ of $t$ is minimal. The term $t$ represents a numerical expression whose value has to be minimized.
%     This is an extension to the modelexpand inference.
%     The complexity of deciding that %an atom  is true in an optimal model is  $\mathbf{\Delta^p_2}$. 
%     a certain $t^\I$ is minimal, is in $\mathbf{\Delta_2^P}$.
%   \item[Propagate($\T,\pS$):]  input:  theory ${\T}$ and partial structure $\pS$; output: the most precise partial structure $\pS_r$ such that for every model $\I\geqp\pS$ of ${\T}$ it is true that $I\geq_p \pS_r$.
%   The complexity of deciding that a partial structure $\mc{S}'$ is  $\mc{S}_r$ is in $\mathbf{\Delta_2^P}$. Note that we assume that all partial structures are functionally consistent, which implies that we also propagate functional integrity 
%   constraints.
%   \item[Deduction] 
%     Input are 2 logical theories $\T_1$  and  $\T_2$. 
%     Output is \textbf{true} if $\T_1$ subsumes $\T_2$ and \textbf{false} otherwise.
%     Or formally, we return:
%     $$\T_2
\item \textbf{Query($\pS,E$):} Input: a (partial) structure $\pS$ and a set expression
  $E=\{\overline{x}\mid \varphi(\overline{x})\}$. %
  Output: the set $A_Q=\{\overline{x}\mid\varphi(\overline x)^\pS=\Tr\}$.
  %Complexity of deciding that a set $A$ is $A_Q$ is in    \textbf{P}.
\item \textbf{Progression($\T,\pS_i$):} In~\cite{iclp/Bogaerts14}, LTC theories (Linear Time Calculus) are proposed, a syntactic subclass of \fodot theories that allow to naturally model dynamic systems. 
 An LTC theory consists of three types of constraints: constraints about the initial situation, %($P(0)$)
invariants, %$(\forall t: P(t) \lor Q(t))$
and ``bistate'' formulas that relate the state on the current point in time with that of the next. %$(\forall t: P(t) \Rightarrow P(t+1))$ 
Note that the specification presented in Subsection \ref{sec:spec} below is an LTC theory.

The \emph{Progression inference}: Input: an LTC theory $T$ and a structure $\pS_i$ that provides information about the state of the system on a time point $t$. Output: a structure  $\pS_{t+1}$ that represents the next state (or a next possible state) at time point $t+1$.  Repeating this process, we can compute all subsequent states, effectively simulating the dynamic system defined by $\T$. 
\end{itemize}

\section{Delegation revocation in the KB paradigm}
\label{sec:KBrs}

In this section, we explain how the KB paradigm can be applied to delegation revocation. For this purpose, we show how the delegation revocation framework defined in Section \ref{sec:rs} can be specified in \fodot, and how inferences on this specification can solve various tasks that arise in the domain. Some of these tasks are tasks that any system implementing the delegation revocation framework needs to solve, while others are tasks that support a user of such a system. 

We have built a prototype in IDP in which this application of the KB paradigm is realized. This prototype also covers the negative authorizations and negative revocation schemes that this paper does not explain due to space restrictions. 
This prototype can be downloaded at \url{http://icr.uni.lu/mcramer/downloads/hagstrom-RDS.zip} and run in IDP 3.

% In this section we sketch our implementation\footnote{The implementation can be downloaded at \url{http://icr.uni.lu/mcramer/index.php?id=3}.} of the eight revocation schemes and the undoing operation in IDP. Because of the nature of IDP, whose inductive definitions suit the recursive character of the revocation schemes very well, the revocation schemes could be implemented in a very straightforward way. The implementation sheds light on both the formal properties and the practical implications of the revocation schemes, and can thus support a developer of an access control system in her decisions concerning the precise nature of the revocation schemes to be included in the system.

\subsection{The \fodot specification of the delegation revocation framework}
\label{sec:spec}

In this subsection, we describe how Hagstr\"om et al.'s delegation revocation framework, which we defined semi-formally in Section \ref{sec:rs}, can be formally specified in \fodot. The full specification can be found in Appendix A.
%the file \texttt{MainTheory.idp} of the prototype that can be downloaded at \url{http://icr.uni.lu/mcramer/downloads/hagstrom-R&DS.zip}.\footnote{For the convenience of the reviewers, we have copied the specification \texttt{MainTheory.idp} into Appendix A, including the comment that clarify various details.}
% Appendix A.\footnote{For a version of the specification that includes the formal specification of the vocabulary, that also covers negative authorizations, and that contains comments that clarify various details, see the file \texttt{MainTheory.idp} of the prototype that can be downloaded at \url{http://icr.uni.lu/mcramer/downloads/hagstrom-RDS.zip}.}  
Both in this subsection and in the appendix, we use IDP syntax for \fodot: The symbols \texttt{\&}, \texttt{|}, \texttt{\~}, \texttt{!} and \texttt{?} mean $\land$, $\lor$, $\neg$, $\forall$ and $\exists$ respectively, and the symbol sequence \texttt{<-} means $\leftarrow$ (in inductive definitions).

% In the description of the \fodot specification, we concentrate on the four deletion schemes. The negative schemes work in a similar way, just that authorizations get inactivated instead of deleted.

The \fodot specification models the change of the authorization specification over time. Principals are modelled as objects of the theory's domain, whereas positive authorizations are modelled by the predicate \texttt{pos\_auth}.
%and \texttt{neg\_auth} for negative authorizations. 
The authorizations cannot be modelled as objects, because they change over time, while \fodot assumes a constant domain of objects. 

We allow for four types of objects: \texttt{Time}, \texttt{principal}, \texttt{scheme} and \texttt{permission}. 
Time points are integers. There is a constant \texttt{SOA} of type \texttt{principal} that denotes the source of authority. The type \texttt{scheme} consists of the four delete revocation schemes (\texttt{WLD}, \texttt{WGD}, \texttt{SLD} and \texttt{SGD}), the four negative revocation schemes (\texttt{WLN}, \texttt{WGN}, \texttt{SLN}, \texttt{SGN}) 
and four schemes for granting the four different kinds of permissions (\texttt{grantTT}, \texttt{grantTF}, \texttt{grantFT} and \texttt{grantFF}). The permissions are \texttt{TT}, \texttt{TF}, \texttt{FT} and \texttt{FF}.
The sign is not part of the permission, because the information provided by the sign is included in the choice between \texttt{pos\_auth} and \texttt{neg\_auth}. 

As IDP only works with finite domains, the type \texttt{Time} actually just consists of a finite set of consecutive integers. There is a constant \texttt{Start} for the first time point, i.e.\ the smallest integer in the domain of time points in a given structure. The unary partial function \texttt{Next} maps a time point \texttt{t} to the next time point \texttt{t+1}, as long as \texttt{t} is not the last time point included in the domain.

The predicate \texttt{pos\_auth} for positive authorizations takes four arguments: \texttt{pos\_auth(t,i,j,a)} means that at time \texttt{t}, a positive authorization from principal \texttt{i} to principal \texttt{j} for permission \texttt{a} is in place. 
The predicate \texttt{neg\_auth}, on the other hand, only takes three arguments, as no permssion needs to be specified for a negative authorization: \texttt{neg\_auth(t,i,j)} means that at time \texttt{t}, a negative authorization from principal \texttt{i} to principal \texttt{j} is in place. 
% There is a tertiary predicate \texttt{pos\_{}auth\_{}start} for specifying the positive authorizations that are in place at the first time point \texttt{Start}.
There are two predicates, the tertiary \texttt{pos\_{}auth\_{}start} and the binary \texttt{neg\_{}auth\_{}start}, for specifying the positive and negative authorizations that are in place at the first time point \texttt{Start}.

Changes in the authorization specification are always triggered by some action by a principal: \texttt{action(t,s,i,j)} means that at time \texttt{t}, principal \texttt{i} performs an action of the (revocation or grant) scheme \texttt{s} affecting principal \texttt{j}. In the case of delete revocations and grants, these actions can lead to authorizations being deleted and/or new authorizations being included in the authorization specification. \texttt{delete(t,i,j,a)} means that between time points \texttt{t-1} and \texttt{t}, the positive authorization from \texttt{i} to \texttt{j} for permission \texttt{a} gets deleted. \texttt{new(t,i,j,a)} means that between time points \texttt{t-1} and \texttt{t}, a new positive authorization from \texttt{i} to \texttt{j} for permission \texttt{a} gets added to the authorization specification.

\texttt{pos\_{}auth} is defined inductively by setting its values at the \linebreak first time point \texttt{Start} to the start configuration specified by \linebreak \texttt{pos\_{}auth\_{}start}, and by modifying its values between time \texttt{t} and \texttt{t+1} according to the changes specified by \texttt{delete} and \texttt{new}:
%[caption={The inductive definition of \texttt{pos\_{}auth}},label={pos_auth}]
\begin{lstlisting}
{pos_auth(Start,p1,p2,a) <- 
	pos_auth_start(p1,p2,a). 
 pos_auth(Next(t),p1,p2,a) <-
	pos_auth(t,p1,p2,a) & 
	~delete(Next(t),p1,p2,a). 
 pos_auth(Next(t),p1,p2,a) <-
	new(Next(t),p1,p2,a).}
\end{lstlisting}
% Since in this sketch of the implementation we are leaving out the negative revocation schemes, we can ignore negative authorizations. In the actual implementation, there are predicates \texttt{FF\_{}delete} and \texttt{new\_{}FF} that specify changes on the negative authorizations, and \texttt{FF} is defined in a way analogous to \texttt{pos\_{}auth} using these change predicated instead of \texttt{delete} and \texttt{new}.

Additionally, there are predicates \texttt{FF\_{}delete} and \texttt{new\_{}FF} that specify changes on the negative authorizations, and \texttt{FF} is defined in a way analogous to \texttt{pos\_{}auth} using these change predicated instead of \texttt{delete} and \texttt{new}.

The predicate \texttt{chain(t,i,a)} expresses that at time \texttt{t}, there exists a rooted delegation chain for principal \texttt{i} with respect to permission \texttt{a}. In Section \ref{sec:rs}, rooted delegation chains are defined by quantifying over sequences of principals. This is in effect a second-order quantification, which is not possible in the first-order language \fodot. However, \texttt{chain(t,i,a)} can be equivalently defined through an inductive definition as follows:\footnote{Note that first-order logic with inductive definitions has an expressivity that lies strictly between the expressivity of first-order and second-order logic.}
%[caption={The definition of \texttt{chain}},label={chain}]
\begin{lstlisting}
{chain(t,SOA,TT).
 chain(t,p1,a1) <- 
	?p2: chain(t,p2,a) &
	pos_auth(t,p2,p1,a1) & 
	 R(a,a1).
 chain(t,p,a) <- 
	chain(t,p,a1) &
	Stronger(a1,a). }
\end{lstlisting}

The predicate \texttt{can\_{}grant(t,i,a)} expresses that principal \texttt{i} has a permission that entitles him to grant the permission \texttt{a}. The predicate \texttt{ind(t,i,j,a)} models the independence of principal \texttt{i} from principal \texttt{j} with respect to a permission \texttt{a}, and the access right of principal \texttt{i}. These two predicates are defined as follows:
%[caption={The definitions of \texttt{can\_{}grant} and \texttt{ind}},label={auxiliary}]
\begin{lstlisting}
{can_grant(t,i,a) <-
	?a1: chain(t,i,a1) & R(a1,a).}
{ind(t,SOA,p,a) <- 
	~ SOA = p.
 ind(t,p1,p2,a) <- 
	~ p1 = p2 
	& ?p: ind(t,p,p2,a1) &
	pos_auth(t,p,p1,a) &
	(a=TT | a=TF) & 
	R(a1,a).}
\end{lstlisting}

Analogously to \texttt{chain} and \texttt{can\_{}grant}, there are predicates \linebreak \texttt{active\_{}chain} and \texttt{can\_{}actively\_{}grant} that additionally take care of the authorizations in the chain being active. 

The different effects of the different deletion revocation schemes are captured by the definitions of the predicates \texttt{delete}, \texttt{inactive} and \texttt{new}. \texttt{delete} is defined via an inductive definition with four clauses:
%[caption={The definition of \texttt{delete} captures the dominance dimension},label={delete}]
\begin{lstlisting}
{delete(Next(t),i,j,a) <- 
	pos_auth(t,i,j,a) &
	action(t,s,i,j) & 
	(s=WLD | s=SLD | s=WGD | s=SGD).
 delete(Next(t),i,j,a) <-
	pos_auth(t,i,j,a) & 
	~ can_grant(Next(t),i,a).      
 delete(Next(t),k,j,a) <-
	pos_auth(t,k,j,a) &
	action(t,SLD,i,j) &  
	?a1: (ind(t,k,i,a1)& R(a1,a)).
 delete(Next(t),z,w,a) <-
	pos_auth(t,z,w,a) & 
	action(t,SGD,i,j) & 
	delete(Next(t),p,w,a1) & 
	~ ?a2: (ind(t,z,i,a2) & 
	R(a2,a)).}
\end{lstlisting}
The first clause just states that in any deletion revocation scheme from $i$ to $j$, the positive authorization from $i$ to $j$ is deleted. The second clause defines the propagation of deletion by specifying that any positive authorization from $i$ to $j$ gets deleted if $i$ is losing its delegation right. The last two clauses capture the meaning of \emph{strong} vs. \emph{weak} dominance by specifying the additional deletions that are needed in strong revocation schemes. 

The predicate \texttt{inactive} specifies which authorizations are inactive. The conditions for inactivating authorizations are analogous to the conditions for deleting authorization, only that they come into play in negative revocations rather than in delete revocations:
\begin{lstlisting}
{inactive(Next(t),i,j,a) <- 
	action(t,s,i,j) &
	pos_perm(Next(t),i,j,a) &
	(s=WLN | s=SLN | s=WGN | s=SGN).
 inactive(t,j,k,a) <- 
	pos_perm(t,j,k,a) &
	~ can_actively_grant(t,j,a).
 inactive(Next(t),z,j,a) <-
	?i:action(t,rs,i,j) & 
	(rs=SLN | rs=SGN) &
	pos_perm(t,z,j,a) &
	~ ?a1: (ind(t,z,i,a1)& R(a1,a)).
 inactive(Next(t),z,w,a) <- 
	action(t,SGN,i,j) &
	inactive(Next(t),p,w,a1) & 
	~inactive(t,p,w,a1)&
	pos_perm(t,z,w,a) &
	~ ?a2: (ind(t,z,i,a2)& R(a2,a)). 
 inactive(Next(t),i,j,a) <- 
	inactive(t,i,j,a) &
	pos_perm(Next(t),i,j,a). }
\end{lstlisting}
The last clause of \texttt{inactive} specifies that an inactivated authorization stays inactive. If an action of undoing a negative revocation was included in the framework, this clause would have to be modified so as to allow for reactivation. But since Hagstr\"om et al. do not define such an undoing of negative revocations, we do not include it in our formal specification of their framework either.

The meaning of \emph{local} vs. \emph{global} propagation is captured by the inductive definition of the predicate \texttt{new}. It consists of four clauses which take care of four different reasons for adding new authorizations. For simplicity, we only present two of these clauses here: 
%[caption={The definition of \texttt{new} captures the propagation dimension},label={new}]
\begin{lstlisting}
new(Next(t),i,j,a) <-  
	? ds:(ds=grantTT | ds=grantTF | 
		ds=grantFT | ds=grantFF) & 
	action(t,ds,i,j) & 
	Relation(a,ds)&
	can_actively_grant(t,i,a) & 
	~ ? a1: (can_actively_grant(t,i,a1) & 
		Relation(a1,ds) & Stronger(a1,a)). 
new(Next(t),i,k,a) <- 
	?j s:(s=WLD | s=SLD) & 
	action(t,s,i,j) & 
	(? z: pos_auth(t,z,k,a) & 
	      can_grant(t,z,a)) & 
	~(? z: pos_auth(t,z,k,a) & 
	      can_grant(Next(t),z,a)). 
\end{lstlisting}
The first clause ensures that new positive authorizations are added when a granting action is performed by a principal.\footnote{The last tree conditions in this first clause ensure that when \texttt{i} cannot grant permission \texttt{a}, the strongest permission that \texttt{i} can grant and that is still weaker than \texttt{a} will actually be granted.} The second clause takes care of adding the authorizations that are added in the last item of each of the definitions of Weak Local Delete and Strong Local Delete in Section \ref{sec:rs}. Informally, this clause says that if in a local revocation scheme revoking a positive authorization from principal \texttt{i} to principal \texttt{j}, \texttt{j} is losing its delegation right, then every positive authorization from \texttt{j} to another principal \texttt{k} must be replaced by a positive authorization of the same authorization type from \texttt{i} to \texttt{k}. This new authorization from \texttt{i} to \texttt{k} ensure that the propagation defined in the second clause of the definition of \texttt{delete} does not continue beyond \texttt{j}.

% Note that we needed to specify the additional strength of the deletion separately for strong local deletes and strong global deletes: This is because we wanted -- in line with the definition of the strong global delete in \cite{Hagstrom} and \cite{Aucher} -- a strong global delete from $i$ to $j$ not only to be strong in the sense of deleting other permissions to $j$ dependent on $i$, but also to delete other permissions dependent on $i$ to descendants of $j$. We doubt, however, whether this additionally strength of the strong global delete would actually be desirable in a real access control system: Strong revocation schemes are usually applied to distrusted principals, whose rights one wants to restrict as much as possible. But there is no reason why another principal, who has a rooted delegation chain independent of this distrusted principal, should have his rights removed only because he also has a rooted delegation chain dependent on the distrusted principal. The version of the strong global delete that we judge more reasonable is the one in which the fourth line of the inductive definition of \texttt{delete} is removed and the third line is also applied to the strong global delete.

% This discussion of the details of the strength of the strong global delete illustrates how modelling revocation schemes in IDP can shed light on the properties of the revocation schemes in a way that can support a developer of an access control system in fixing the specification of the schemes to be implemented in the system.

The predicate \texttt{access\_{}right(t,i)} means that principal \texttt{i} has access right at time \texttt{t}: 
%[caption={The definition of \texttt{new} captures the propagation dimension},label={new}]
\begin{lstlisting}
{access_right(t,p) <-
	active_chain(t,p,a).
access_right(t,p) <-
	?p1:can_actively_grant(t,p1,a)& 
	pos_auth(t,p1,p,a) & 
	~inactive(t,p1,p,a).}
\end{lstlisting}

\subsection{Using inferences to solve various tasks}
In this subsection we explain how different logical inferences, when applied to the \fodot specification explained above, can solve various tasks that can be useful both for implementing a system which allows for delegation revocation and for supporting a user of such a system. 

Let us first consider tasks that a system that implements Hagstr\"om et al.'s delegation revocation framework needs to solve. Given a certain state of the system, defined by which authorizations are currently included in the authorization specification, and a given action (a delegation or revocation performed by some principal), the new state of the system after this action needs to be determined. This task can be performed using the Modelexpand inference as follows: Let $T$ be the \fodot specification of the delegation revocation framework. Let $\pS$ be a partial structure with the following properties:
\begin{itemize}
 \item The time domain of $\pS$ contains only the two time points $0$ and $1$.
 \item $\pS$ assigns to the predicate \texttt{pos\_{}auth\_{}start} the set of all authorizations currently included in the authorization specification.
 \item $\pS$ assigns to the predicate \texttt{action} the given action at time $0$.
 \item The value of all other predicates is undefined in $\pS$.
\end{itemize}
In this case, Modelexpand($T$,$\pS$) is a total structure that expands $\pS$ and that is a model of $\T$. Being a total structure, it assigns to \texttt{pos\_{}auth} a set $A$ of quintuples of the form $(t,i,j,\langle\mathit{sign}\rangle,a)$, where $t$ is a time point ($0$ or $1$), $i$ and $j$ are principals, $\langle\mathit{sign}\rangle$ is a sign, and $a$ is a permission. Then the set $A' := \{(i,j,\langle\mathit{sign}\rangle,a) \;|\; (1,i,j,\langle\mathit{sign}\rangle,a) \in A \}$ is the set of authorizations that constitutes the authorization specification after the action. (Note that since all predicates other than \texttt{pos\_{}auth\_{}start} and \texttt{action} are defined in $T$ through an inductive definition, there is a unique model of $T$ that expands $\pS$, so that the result of this inference is deterministic.)

This way we can determine the effect of a single action. It could be iterated by setting the value of \texttt{pos\_{}auth\_{}start} to be $A'$ for the next iteration of this procedure. But the IDP system also supports an inference, namely the Progression inference, that is designed for this kind of temporal progression of a structure based on a theory with a type for time. The input  structure of this inference provides information about the state of the system on a time point $t$; in our case that is the authorization specification at a given time. The output is a structure that represents the state of the authorization specification at time point $t+1$. So the step of extracting $A'$ from $A$ that was required for iterating the above inference is no longer needed. So this inference can more straightforwardly be iterated, giving rise to an interactive simulation of the  progression of the system state through time.

Of course, a system implementing Hagstr\"om et al.'s framework does not only need to determine how the authorization specification changes over time, but also needs to determine whether a principal requesting access or performing a certain administrative action actually has access right or the right to perform the action in question. This can be done with the Query inference: For example, if $\pS$ is the partial structure that assigns to the predicate \texttt{pos\_{}auth\_{}start} the set of all authorizations currently included in the authorization specification, Query($\pS$,$\{\texttt{i} \; | \; \texttt{access\_{}right(Start,i)} \}$) returns the set of principals that have access right according to the current authorization specification.

%TODO: Update with LTC (2014) paper, which has the inferences isinvariant/2 and isinvariant/3 (but they seem to be buggy at the moment).
When designing a system, one can avoid an erroneous design by specifying invariants that the system must satisfy at any moment during the execution of the system, and verify that these invariants are actually satisfied by the system. In the case of a system based on Hagstr\"om et al.'s delegation revocation system, examples of invariants that the system must satisfy are the connectivity property  and the active-connectivity property defined in subsections \ref{delegation} and \ref{negative}. We must, of course assume, that the system starts in a state that satisfies these two  properties. All that remains to be shown, then, is that if the connectivity property holds at some time point $t$, it must also hold at the next time point $t+1$, and similarly for the active-connectivity property.

One way that this can be done is by calling an automated theorem prover to prove this implication. However, this is not always viable, as the theory may be too complex for an automated theorem prover to be able to find a proof of the invariant. This is the case for our specification. 

Another possibility is to prove that the invariant holds in fixed structures. In our case, we can fix a partial structure $\pS$ with time points $0$ and $1$ and without any information about the predicates. In this case, the only information that we are fixing is the number of principals. We can then prove that the invariant holds for a fixed number of principals by establishing, using Modelexpand($T'$,$\pS$), that there is no total structure expanding $\pS$ that is a model of the theory $T'$ consisting of our specification $T$ together with the statement that the connectivity property holds at time point $0$ but not at time point $1$. With this method, we have verified the connectivity property for any system with $n$ principals for $n \leq 8$.

Despite this limitation to very small domains, this limited verification can be useful for avoiding erroneous design, as errors tend to already show up at relatively small domains. It should be added that the logical methodology of the KB paradigm lends itself well to the usage of interactive theorem provers common in software verification in order to fully verify invariants over complex specifications. The integration of IDP and interactive theorem provers is, however, still future work.

Finally, let us turn to a task that supports a user of a system based on Hagstr\"om et al.'s delegation revocation framework: A principal $i$ may want to reach a certain outcome, e.g.\ that a given principal $j$  should no longer have a certain right while another principal $k$ is unaffected. $i$ may want to find out all revocation schemes that lead to the desired outcome. This can be achieved by computing Allmodels($T_2$,$\pS$), where $\pS$ is the structure that assigns the current authorization specification to \texttt{pos\_{}auth\_{}start}, and $T_2$ is the theory consisting of our specification of the delegation revocation framework together with the statement that the action at time point $0$ is performed by $i$, and the statement that the desired outcome holds at time point $1$. The values of the predicate \texttt{action} at time point $0$ in the models returned by Allmodels($T_2$,$\pS$) are the actions that $i$ can perform in order to get the desired outcome.

Furthermore, $i$ may want to reach a certain outcome for some given principals while minimally influencing the permissions of other principals. In that case, $i$ can define a cost function, e.g.\ that every change in a permission of a principal has a cost of $1$, and search the models returned by Allmodels($T_2$,$\pS$) for the one with the minimal cost.

The IDP prototype that we have built can perform all the different tasks described in this subsection.

\section{Related work}
\label{sec:related}
% \subsection{Other applications of the KB paradigm}
While the KB paradigm and its implementation \idp, are fairly young, its applicability has been investigated and illustrated in multiple domains. In \cite{iclp/VanHertum13}, the connection with Business Rules was investigated. Business Rules are well-represented in industry for knowledge-intensive applications and as such were used as a comparison to evaluate the KB paradigm. A typical Business Rules application, the EU-Rent Car-Rental company, was modelled in \fodot, and two use cases were investigated. 

In \cite{ppdp/VlaeminckVD09}, the authors looked at applications of the \idp system, for interactive configuration systems, where the system is used to guide a user through a search space, looking for a valid configuration. The advantages of an explicit modeling of domain knowledge in configuration were large: the adaptability in case the domain knowledge changes and the fact that the same specification of knowledge could be reused in different tasks being the most important. This work was extended in \cite{padl/VanHertumDJD16}, where eight different reasoning tasks used in a configuration system were identified and implemented using logical inferences on 1 knowledge base, containing all domain knowledge.

% \subsection{Other logical approaches to delegation revocation} 
While the KB paradigm has not been previously applied to the problem of delegation revocation, other logical methods have been applied to this access control problem: Aucher et al.\ \cite{Aucher} presented a formalization of Hagstr\"om et al.'s eight revocation schemes in a dynamic variant of propositional logic that resembles imperative programming languages. Furthermore, they extended their formalization with a notion of trust. Unlike our specification, which specifies the eight revocation schemes by specifying the formal properties of the three dimensions of the classification, their formalization defines each of the eight revocation schemes separately. Their formalization only supports the tasks of determining the state of the system after a certain action and of determining whether a user has a certain permission given a certain state of the system; the other tasks described in this paper are not supported by their formalization.

The main author of the current paper has defined a modified version of Hagstr\"om et al.'s delegation revocation framework \cite{postulates} as well as \emph{Trust Delegation Logic}, a logic of trust designed for studying the reasons for performing different revocation schemes defined \cite{Cramer}. This work was motivated by problems we encountered with Hagstr\"om et al.'s delegation revocation framework when we produced the first \fodot specification of the framework; these problems are documented in \cite{postulates}, \cite{Cramer} and \cite{Cramer16}. In general, it should be noted that formally specifying something in \fodot can help understanding it better and uncovering problematic features. As a further example of this, our work on \fodot specifications of delegation revocation frameworks has also shown us that Aucher et al.'s \cite{Aucher} formalization actually deviates from Hagstr\"om et al.'s delegation revocation framework in multiple respects (see \cite{Cramer16} for details).

\section{Conclusion}
\label{sec:conclusion}
In this paper, we have explained the benefits of the knowledge base paradigm when applied to delegation revocation.
The knowledge base paradigm proposes a strict separation between knowledge and problem solving.
In our application, the knowledge is represented by an \fodot specification of Hagstr\"om et al.\ \cite{Hagstrom} delegation revocation framework. 
By applying various logical inferences to this specification, multiple tasks that arise when implementing or using a delegation revocation system were solved. 
This way, the same information was reused for solving various problems.

Our work constitutes a proof of concept, and we hope that it will inspire other researchers in computer security to consider the possibility of applying the methodology of the knowledge base paradigm to their research.

\bibliographystyle{IEEEtran}
\bibliography{references,idp-latex/krrlib}

\appendix

\section{The \foid specification of the delegation revocation framework}

\begin{lstlisting}
{	chain(t,SOA,TT).
        
	chain(t,p1,a1) <- ?p2: chain(t,p2,a) &
        	pos_perm(t,p2,p1,a1) & 
        	R(a,a1).
        
        chain(t,p,a) <- chain(t,p,a1) & 
        	Stronger(a1,a). }
	
{	active_chain(t,SOA,TT).
        
	active_chain(t,p1,a1) <- ?p2: active_chain(t,p2,a) &
        	pos_perm(t,p2,p1,a1)  & 
        	R(a,a1) & 
        	~inactive(t,p2,p1,a1).
        
        active_chain(t,p,a) <- active_chain(t,p,a1) & 
        	Stronger(a1,a). }
    
{	can_grant(t,i,a) <- ?a1: chain(t,i,a1) & 
        	R(a1,a). }
    
{	can_actively_grant(t,i,a) <- ?a1: active_chain(t,i,a1) & 
        	R(a1,a). }
	 
{	ind(t,SOA,p,a) <- SOA ~= p.
        ind(t,p1,p2,a) <- p1 ~= p2 & 
        	?p: ind(t,p,p2,a1) &
        	pos_perm(t,p,p1,a) & 
        	(a=TT | a=TF) & 
        	R(a1,a). }
	  
{	access_right(t,p) <- active_chain(t,p,a).

	access_right(t,p) <- ?p1:can_actively_grant(t,p1,a) & 
        	pos_perm(t,p1,p,a) & 
        	~inactive(t,p1,p,a). }
    
//Here we define which authorizations to delete under the different delete revocation schemes
{	delete(Next(t),i,j,a)<- pos_perm(t,i,j,a) &
        	action(t,s,i,j) &
		(s=WLD | s=SLD | s=WGD | s=SGD).
        
	//Define deletion via rooted delegation chains:
	delete(Next(t),j,k,a)<-pos_perm(t,j,k,a) &
		~ can_grant(Next(t),j,a).//We don't use can_actively_grant here, because even an inactive delegation chain can prevent deletion
        
	//Extra deletion rule for Strong Local Delete scheme:
	delete(Next(t),k,j,a)<-action(t,SLD,i,j) &
		pos_perm(t,k,j,a) &
		~ ?a1: (ind(t,k,i,a1)& R(a1,a)).
        
	//Extra deletion rule for Strong Global Delete scheme:
	delete(Next(t),z,w,a)<-action(t,SGD,i,j) &
		delete(Next(t),p,w,a1) &
		pos_perm(t,z,w,a) &
		~ ?a2: (ind(t,z,i,a2)& R(a2,a)). }

{	// A delegation scheme leads to a new positive authorization if the principal performing it has the right to grant such an authorization:
        new(Next(t),i,j,a) <-  ? ds:(ds=grantTT | ds=grantTF | ds=grantFT | ds=grantFF) & 
        	action(t,ds,i,j) & 
        	// The following three conditions ensure that when i cannot grant permission a, the strongest permission that i can grant and that is still weaker than a will actually be granted:
        	Relation(a,ds)&
        	can_actively_grant(t,i,a) & 
 		~ ? a1: (can_actively_grant(t,i,a1) & Relation(a1,ds) & Stronger(a1,a)). 
        
	// In local revocation schemes, if an authorization from i to z is revoked and an authorization from z to another node k is getting deleted, there should be a new authorization from i to k.
	new(Next(t),i,k,a) <- ?j s:(s=WLD | s=SLD  | s=WLN | s=SLN) & 
		action(t,s,i,j) & 
        	//check that there was an active authorization from z to k authorization was active:
		(? z: pos_perm(t,z,k,a) & can_actively_grant(t,z,a)) & 
        	//and that the authorization is is about to be deleted or inactivated:
		~(? z: pos_perm(t,z,k,a) & can_actively_grant(Next(t),z,a)). 

        // An authorization from j to k that gets deleted because j loses her right to grant it gets replaced by the strongest authorization that j can grant and that is weaker than the deleted authorization.
        new(Next(t),j,k,a1) <- pos_perm(t,j,k,a) &
		~can_grant(Next(t),j,a) &
        	can_grant(Next(t),j,a1) &
        	Stronger(a,a1) &
        	// a1 is the strongest possible authorization that j can grant:
        	~ ? a2: (can_grant(Next(t),j,a2) & Stronger(a,a2) & Stronger(a2,a1)).

	// The same as above but for inactivation rather than deletion of the authorization from j to k:
	new(Next(t),j,k,a1) <- ?i RS:(RS=WLN | RS=WGN) & 
		action(t,RS,i,j) & 
        	pos_perm(t,j,k,a) &
		can_actively_grant(Next(t),j,a) &
        	can_actively_grant(Next(t),j,a1) &
        	Stronger(a,a1) &
        	// a1 is the strongest possible authorization that j can grant.
        	~ ? a2: (can_actively_grant(Next(t),j,a2) & Stronger(a,a2) & Stronger(a2,a1)). }
     	
// Here we define which links to inactivate under the different negative revocation schemes. The definitions are analogous to the definitions of the delete predicate for delete revocation schemes.
{	inactive(Next(t),i,j,a)<-action(t,s,i,j)&
        	pos_perm(Next(t),i,j,a) &
		(s=WLN | s=SLN | s=WGN | s=SGN).
			
	inactive(t,j,k,a)<-pos_perm(t,j,k,a) &
        	~ can_actively_grant(t,j,a).
        
       inactive(Next(t),z,j,a)<-?i:action(t,rs,i,j) & 
        	(rs=SLN | rs=SGN) &
		pos_perm(t,z,j,a) &
		~ ?a1: (ind(t,z,i,a1)& R(a1,a)).
        
       inactive(Next(t),z,w,a)<-action(t,SGN,i,j) &
		inactive(Next(t),p,w,a1) & ~inactive(t,p,w,a1)&
		pos_perm(t,z,w,a) &
		~ ?a2: (ind(t,z,i,a2)& R(a2,a)). 
	
	inactive(Next(t),i,j,a) <- inactive(t,i,j,a) &
        	pos_perm(Next(t),i,j,a). }
    
// When a negative revocation scheme is granted from i to j, a negative permission from i to j is created:
{	new_neg_perm(Next(t),i,j)<-action(t,RS,i,j)&
        	(RS = WLN | RS = SLN | RS=WGN | RS=SGN). }

{	pos_perm(Start,p1, p2,a)<-pos_perm_start(p1, p2,a). 
	pos_perm(Next(t), p1, p2,a)<-pos_perm(t, p1, p2,a) & ~delete(Next(t), p1, p2,a). 
	pos_perm(Next(t), p1, p2,a)<-new(Next(t), p1, p2,a). }
    
{	neg_perm(Start,p1,p2)<-neg_perm_start(p1,p2).//initiating neg-perm
	neg_perm(Next(t),p1,p2)<-neg_perm(t,p1,p2).
	neg_perm(Next(t),p1,p2)<-new_neg_perm(Next(t),p1,p2). }
     
    {R(TT,TT). R(TF,TF). R(TT,TF). R(TT,FT). R(TF,FF).R(TT,FF).}
    
    {Stronger(TT,TF).Stronger(TT,FT).Stronger(TT,FF). Stronger(TF,FF). Stronger(FT,FF). }
    
    { Relation(TT,grantTT). Relation(TF,grantTT). Relation(FT,grantTT). Relation(FF,grantTT). Relation(TF,grantTF). Relation(FF,grantTF). Relation(FT,grantFT). Relation(FF,grantFF). }

//At most one revocation scheme at every Time point.	 
?<2 s p1 p2: action(t,s,p1,p2).
\end{lstlisting}

\end{document}